\documentclass[a4paper]{ifacconf}

\usepackage{listings}
\usepackage{amsmath,amssymb,pstricks,color,dsfont}
\usepackage{savesym}
\savesymbol{AND}
\usepackage{natbib}
\usepackage{graphicx}
\usepackage{psfrag,graphicx,epsfig}
\usepackage{algorithm,xspace}
\usepackage{subfig}
\usepackage{float}
\usepackage{graphicx}
\usepackage{epstopdf}
\usepackage{url}
\usepackage{algcompatible}
\usepackage{setspace}
\renewcommand{\algorithmicrequire}{\textbf{\textit{Input: }}}
\renewcommand{\algorithmicensure}{\textbf{\textit{Return: }}}

\newcommand{\nnum}{\nonumber}

\newcommand{\EQ}{\begin{eqnarray}}
\newcommand{\EN}{\end{eqnarray}}
\newcommand{\EQQ}{\begin{eqnarray*}}
\newcommand{\ENN}{\end{eqnarray*}}

\DeclareMathOperator{\argmin}{arg\,min}

\newcommand{\vect}[1]{\mathbf{#1}}

\newcommand{\bremark}{\begin{remark} \begin{rm} }
\newcommand{\eremark}{ \end{rm} \rule{1mm}{2mm}
\end{remark} }
\newcommand{\btheorem}{\begin{theorem} \begin{rm} }
\newcommand{\etheorem}{ \end{rm} \rule{1mm}{2mm}
\end{theorem} }
\newcommand{\blemma}{\begin{lemma} \begin{rm} }
\newcommand{\elemma}{ \end{rm} \rule{1mm}{2mm}
\end{lemma} }
\newcommand{\bcorollary}{\begin{corollary} \begin{rm} }
\newcommand{\ecorollary}{ \end{rm} \rule{1mm}{2mm}
\end{corollary} }
\newcommand{\bdefinition}{\begin{definition}\begin{rm} }
\newcommand{\edefinition}{ \end{rm} \rule{1mm}{2mm}
\end{definition} }
\newcommand{\bproposition}{\begin{proposition} \begin{rm} }
\newcommand{\eproposition}{ \end{rm} \rule{1mm}{2mm}
\end{proposition} }
\newcommand{\bexample}{\begin{example} \begin{rm} }
\newcommand{\eexample}{ \end{rm} \rule{1mm}{2mm}
\end{example} }
\newcommand{\basm}{\begin{assumption} \begin{rm}}
\newcommand{\easm}{\end{rm} 
\end{assumption}}

\newcommand{\trace}{\operatorname{tr}}

\newtheorem{theorem}{\bf Theorem}[section]
\newtheorem{lemma}{\bf Lemma}[section]
\newtheorem{definition}{\bf Definition}[section]
\newtheorem{remark}{\bf Remark}[section]
\newtheorem{corollary}{\bf Corollary}[section]
\newtheorem{assumption}{\bf Assumption}[section]

\newcommand\oprocendsymbol{\hbox{$\blacksquare$}}
\newcommand\oprocend{\relax\ifmmode\else\unskip\hfill\fi\oprocendsymbol}

\begin{document}
\begin{frontmatter}

\title{
Towards Resilient UAV: Escape Time in GPS Denied Environment with Sensor Drift\thanksref{footnoteinfo}} 


\thanks[footnoteinfo]{This work has been supported by the National Science Foundation (ECCS-1739732 and CMMI-1663460).}

\author[First]{Hyung-Jin Yoon, Wenbin Wan, Hunmin Kim, Naira Hovakimyan}
\author[Second]{Lui Sha},
\author[Third]{Petros G. Voulgaris}

\address[First]{Department of Mechanical Science and Engineering,}
\address[Second]{Department of Computer Science,}
\address[Third]{Department of Aerospace Engineering,\\
University of Illinois at Urbana-Champaign (UIUC), Urbana, IL 61801, USA.
(email: \{hyoon33, wenbinw2, hunmin, nhovakim, lrs, voulgari\}@illinois.edu)}

\begin{abstract}
This paper considers a resilient state estimation  framework for unmanned aerial vehicles (UAVs) that integrates a Kalman filter-like state estimator and an attack detector. When an attack is detected, the state estimator  uses only IMU signals as the GPS signals do not contain legitimate information. This limited sensor availability induces a sensor drift problem questioning the reliability of the sensor estimates. 
We propose a new resilience measure, {\em escape time}, as the safe time within which the estimation errors remain in a tolerable region with  high probability. This paper analyzes the stability of the proposed resilient estimation framework and quantifies a lower bound for the escape time.
Moreover,  simulations of the UAV model demonstrate the performance of the proposed framework and provide analytical results.
\end{abstract}

\begin{keyword}
Resilient estimation, Stochastic system, Unmanned aerial vehicle
\end{keyword}

\end{frontmatter}
\section{Introduction}
Unmanned aerial vehicles (UAVs) have become popular as commercial, industrial and educational platforms. The mechanical simplicity and agile maneuverability appeal to many applications, such as media production, inspection, and precision agriculture. In all these applications, UAVs need reliable state estimation (e.g. position, velocity) to perform various tasks. Most state estimation techniques for UAVs use an inertial measurement unit (IMU) and a global positioning system (GPS) receiver. However, GPS is vulnerable to spoofing attacks as demonstrated in~\cite{warner2003gps}. In~\cite{warner2003gps}, the Vulnerability Assessment Team at Los Alamos National Laboratory demonstrated that GPS spoofing attacks can be easily implemented by civilians using GPS satellite simulator. Furthermore, increasing applications of UAVs extend the area of operation to the urban areas, where GPS signals are weak or denied due to other structures such as skyscrapers, elevated highways, and bridges.

Resilient UAV navigation requires timely attack detection and mitigation. From controls perspective, traditionally the GPS spoofing attack has been modeled as a malicious signal injection. Attack detection research against malicious signal injection  has been studied extensively for the last several years. The attack detection problem was formulated as an $\ell_0$/$\ell_\infty$ optimization problem in~\cite{fawzi2014secure,pajic2014robustness}.
In~\cite{mo2014detecting}, an active detection scheme, by adding random disturbance signal to the optimal control input, was proposed to increase the detection rate trading off for optimality.
In~\cite{mo2010false}, the authors identify maximum deviations of the state due to the sensor attacks, while remaining stealthy due to the detection.

On the other hand,  since the GPS signal injected by the attacker would cause a discrepancy in the raw antenna signal, the GPS spoofing attack can be detected by examining the raw signal received by the antenna. For example, the shape of the GPS signal strength in polar coordinates was used to detect the GPS attack in~\cite{mcmilin2014single}.
In~\cite{chen2013validation}, an array of GPS antenna was used to detect the discrepancy compared to the normal situations. The methods using the raw GPS signals in~\cite{mcmilin2014single, chen2013validation} have the potential to detect the stealthy attacks defined in~\cite{mo2010false}. However, the methods using the raw antenna signals usually require modifications of the hardware or the low-level computing modules.

Unbiased state estimation in  adversarial environments can be challenging, because the estimator accumulates errors due to attacks. In such cases, simple state detection can be a preferred method. Incomplete list of the related results includes~\cite{mo2010false,pajic2014robustness,yong2015resilient,kim2017attack}. In these efforts, the difference between the measured output and the predicted output has been used to detect attacks and exclude corrupted sensor measurements. The rest of the uncompromised redundant sensors are subsequently used for estimation. We depart from this approach and consider limited sensor redundancy. In particular, the UAV model becomes undetectable in GPS-denied environments. GPS denied state estimation has been studied in~\cite{fuke1996dead,chung2001sensor,bevly2007cascaded}, wherein the focus is on establishing the system output matrices that can be used for  standard (error state) Kalman filter for dead reckoning. 

\textbf{Contribution.} This paper proposes a resilient estimation framework for UAVs. The framework consists of an attack detection module and a state estimator that operates in two modes: (1) normal and (2) emergency. In the normal mode, all available sensor signals are used to estimate the state. In the emergency mode, only IMU signals are fed to the state estimator. The limited sensor availability leads to the sensor drift problem, and the estimates become gradually unreliable. We quantify a new resilience measure, the escape time, which is defined by the safe time within which  the estimation errors remain within a tolerable region with a high probability. 
We analyze the stability of the proposed estimator and find a lower bound of the escape time. Simulations are conducted to show the effectiveness of the proposed framework.

The remainder of this paper is organized as follows: In Section~\ref{sec:pre}, we introduce the notation convention in our paper and the dynamic system model. In the same section, we formulate the problem. In Section~\ref{sec:deg}, we propose a resilient state estimation and detection method for GPS attack detection. Section~\ref{sec:esc} presents the stability analysis of the proposed estimator, and studies escape time to avoid instability. The significance of the escape time and the potential impact of it are described in Section~\ref{sec:disc}. In Section~\ref{sec:sim}, a numerical simulation of a UAV under GPS spoofing attack is presented.

\section{Preliminaries}\label{sec:pre}
This section discusses some preliminary  notations/notions, system models, problem formulation, and $\chi^2$ attack detector.

\subsection{Notations}
We use the subscript $k$ of $\vect{x}_k$ to denote the time index; ${\mathbb R}^n$ denotes the n-dimensional Euclidean space; ${\mathbb R}^{n \times m}$ denotes the set of all $n \times m$ real matrices; $\vect{A}^\top$ denotes the transpose of matrix $\vect{A}$; $\vect{I}$ denotes the identity matrix with an appropriate dimension; $\|\cdot\|$ denotes the standard Euclidean norm for vector or an induced matrix norm; ${\mathbb E}[\,\cdot\,]$ denotes the expectation operator; $\times$ is used to denote matrix multiplication when the multiplied terms are in different lines.

\subsection{System model}
We use the following linear model to consider the flight system dynamics and the attacker model:
\begin{equation}
\begin{aligned}
\vect{x}_{k}&= \vect{A} \vect{x}_{k-1}+ \vect{B} \vect{u}_{k-1} + \vect{w}_{k-1}\\
\vect{y}_{k}^{G} & = \vect{C}^G\vect{x}_k   + \vect{d}_k+ \vect{v}_k^{G}\\
\vect{y}_{k}^{I} & = \vect{C}^I(\vect{x}_k-\vect{x}_{k-1})   + \vect{v}_k^I,
\end{aligned}
\label{e000}
\end{equation}
where $\vect{x}_k \in {\mathbb R}^n$, $\vect{y}_k^G \in {\mathbb R}^{m_G}$, $\vect{y}_k^I \in {\mathbb R}^{m_I}$ are the state, the GPS measurement, and the IMU measurement, respectively.
IMU returns a noisy measurement of the state difference. The noise signals $\vect{w}_k$, $\vect{v}_k^G$, $\vect{v}_k^I$ are assumed to be independent and identically distributed (i.i.d.) Gaussian random variables with zero means and  covariances
${\mathbb E}[\vect{w}_k \vect{w}_k^\top]=\vect{\Sigma}_w$,
${\mathbb E}[\vect{v}_k^G (\vect{v}_k^G)^\top]=\vect{\Sigma}_G$,
${\mathbb E}[\vect{v}_k^I (\vect{v}_k^I)^\top]=\vect{\Sigma}_I$, respectively.
The vector $\vect{d}_k \in {\mathbb R}^{m_G}$ is the GPS spoofing attack, which is unknown to the defender. We assume that the attacker can inject any signal $\vect{d}_k$ into $\vect{y}_k^G$.

\subsection{Problem formulation}
Given the system~\eqref{e000} with two sensors GPS and IMU, the defender aims to detect the GPS spoofing attack and resiliently estimate the state. Furthermore, the defender needs to analyze the reliability of the state estimates in the adversarial situation.

\subsection{$\chi^2$ attack detector}
For linear systems in~\eqref{e000} with Gaussian additive noises $\vect{w}_k$, $\vect{v}_k^G$, and $\vect{v}_k^I$,
state estimations of the standard Kalman filter (KF) are Gaussian as well.
Through this observation, the $\chi^2$ statistical test is widely used in attack detection,~\cite{teixeira2010cyber,mo2014detecting,guo2018roboads}, to distinguish whether the error is induced by statistical noises or attacks.
In particular, the $\chi^2$ test has two hypothesis:
\begin{equation}
    H_0: \vect{d}_k=0, \ H_1: \vect{d}_k \neq 0.
    \label{e000.1}
\end{equation}
By testing~\eqref{e000.1}, we interpret the result as the following:
\begin{enumerate}
    \item Rejecting $H_0$ (accepting $H_1$): there is significant evidence that the error is not zero and the error can be due to the attack.
    \item Keeping $H_0$: we do not have enough evidence to believe that there is an attack.
\end{enumerate}

\section{Resilient State Estimation Design}\label{sec:deg}
The proposed estimation and detection system consists of the attack detector and the state estimator. The attack detector performs a statistical hypothesis test to decide whether the GPS signal is being attacked, based on output prediction error and error-covariance estimated by the state estimator. Depending on the result of the hypothesis testing, the state estimation switches its mode between the \textit{normal mode} and the \textit{emergency mode}. In the normal mode, the state estimator uses both GPS and IMU to estimate the state and detect an attack. If an attack is detected, the estimator switches to emergency mode, where IMU is used to estimate the state. If the attack detector determines that the GPS signals are clean, the state estimation can return to the normal mode.
\begin{figure}[thpb]
\centering
 \includegraphics[width=0.48\textwidth]{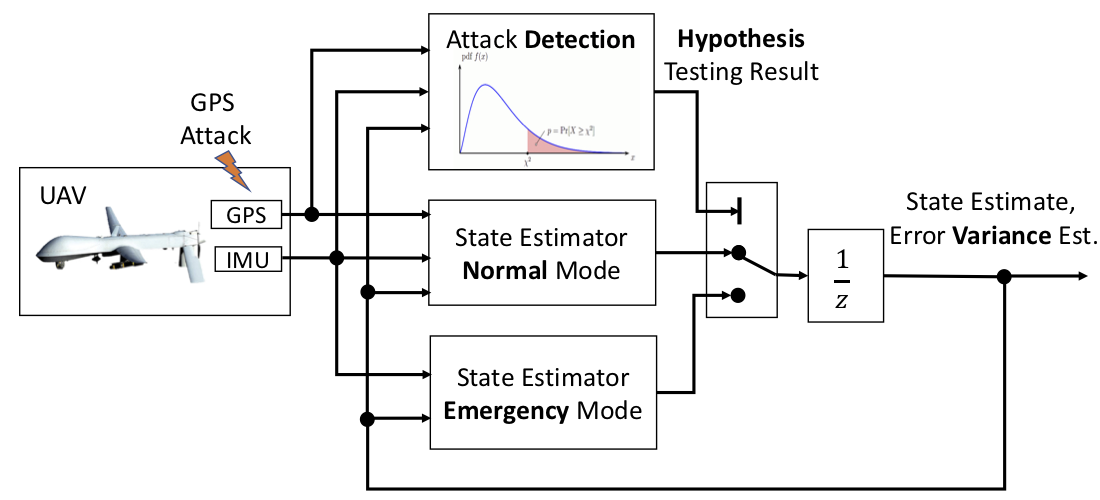}
 \caption{A resilient state estimation framework consisting of GPS attack detection and two modes (normal and emergency) state estimation.}
\medskip
\label{fig:Diagram}
\end{figure}

\subsection{Normal mode}\label{sec:normal-mode}
\textbf{State estimation.}
The defender implements an estimator and $\chi^2$ detector to estimate the state and detect the GPS spoofing attack.
The following KF-like state estimator is used to estimate the current state:
\begin{align*}
\hat{\vect{x}}_{k} &= \vect{A} \hat{\vect{x}}_{k-1}+ \vect{B} \vect{u}_{k-1}+ \vect{K}_k^G (\vect{y}_k^G - \vect{C}^G( \vect{A}\hat{\vect{x}}_{k-1} {+} \vect{B} \vect{u}_{k-1})) \nnum\\
&+ \vect{K}_k^I (\vect{y}_k^I  -\vect{C}^I( \vect{A} \hat{\vect{x}}_{k-1} + \vect{B} \vect{u}_{k-1}-\hat{\vect{x}}_{k-1}))
\end{align*}
\begin{align}
\vect{P}_k&= (\vect{A}-\vect{K}_k\vect{C}\vect{A}+\vect{K}_k\vect{D}\vect{C})\vect{P}_{k-1}(\vect{A}-\vect{K}_k\vect{C}\vect{A}{+}\vect{K}_k\vect{D}\vect{C})\nnum\\
&+(\vect{I}-\vect{K}_k\vect{C})\vect{\Sigma}_w (\vect{I}-\vect{K}_k\vect{C})^\top+\vect{K}_k \vect{\Sigma}_y \vect{K}_k^\top\nnum\\
&\triangleq f(\vect{P}_{k-1}, \vect{K}_k),
\label{eq:P_update}
\end{align} 
where $\vect{K}_k = [\vect{K}_k^G, \ \vect{K}_k^I]$,
\begin{align*}
&\vect{C}=
    \left[
    \begin{array}{c}
    \vect{C}^G\\ \vect{C}^I\\
    \end{array}
    \right], \ 
\vect{\Sigma}_y =
    \left[
    \begin{array}{cc}
    \vect{\Sigma}_G& 0\\
    0&\vect{\Sigma}_I\\
    \end{array}
    \right], \ 
\vect{D} =
    \left[
    \begin{array}{cc}
    0&0\\
    0&\vect{I}\\
    \end{array}
    \right].
\end{align*}
The optimal gain $\vect{K}_k$ can be obtained by solving the following problem:
$
    \min_{\vect{K}_k} \trace{(\vect{P}_k)},
$
which is an unconstrained convex optimization problem.
By taking its derivative with respect to decision variable $\vect{K}_k$ and setting it equal to zero, we have
\begin{align*}
    &(\vect{A}-\vect{K}_k\vect{C}\vect{A}+\vect{K}_k\vect{D}\vect{C})\vect{P}_{k-1}(-\vect{C}\vect{A}+\vect{D}\vect{C})^\top\nnum\\
    &-(\vect{I}-\vect{K}_k\vect{C})\vect{\Sigma}_w \vect{C}^\top + \vect{K}_k \vect{\Sigma}_{y}=0,
\end{align*}
and the solution is
\begin{align}
&\vect{K}_k=(\vect{A}\vect{P}_{k-1}(\vect{C}\vect{A}-\vect{D}\vect{C})^\top+\vect{\Sigma}_w\vect{C}^\top)\nnum\\
&\times\left((\vect{C}\vect{A}-\vect{D}\vect{C})\vect{P}_{k-1}(\vect{C}\vect{A}-\vect{D}\vect{C})^\top\right.+\left.\vect{C}\vect{\Sigma}_w\vect{C}^\top+\vect{\Sigma}_y\right)^{-1}\nnum\\
&\triangleq g(\vect{P}_{k-1}).    
\label{eq:K_law}
\end{align}

\textbf{Attack detection.}
We implement $\chi^2$ statistic test in~\eqref{e000.1} using CUSUM (CUmulative SUM) algorithm, which is widely used in change detection research~\cite{page1954continuous,barnard1959control,lai1995sequential}.

Before proposing a detection algorithm, we consider some properties of attack vector estimates.
Since $\vect{d}_k = \vect{y}_k^G - \vect{C}^G \vect{x}_k - \vect{v}_k^G$, given the previous state estimate $\hat{\vect{x}}_{k-1}$ by the state estimator, we estimate the attack vector by comparing the sensor output and the output prediction:
\begin{align*}
    \hat{\vect{d}}_k     &=\vect{y}_k^G - \vect{C}^G ( \vect{A}\hat{\vect{x}}_{k-1} + \vect{B} \vect{u}_{k-1}).
\end{align*}
The current estimate $\hat{\vect{x}}_k$ should not be used, because it is correlated with the current output; i.e., ${\mathbb E}[\hat{\vect{x}}_k\vect{y}_k^G] \neq 0$.

Due to the Gaussian noises $\vect{w}_k$ and $\vect{v}_k$ injected to the linear system in~\eqref{e000}, the state estimates follow Gaussian distribution, since any finite linear combination of Gaussian distributions is also Gaussian.
Similarly, $\hat{\vect{d}}_k$ is Gaussian as well, and thus the use of $\chi^2$ test~\eqref{e000.1} is justified.
The $\chi^2$ test compares the normalized attack vector estimate $\hat{\vect{d}}_k^\top (\vect{P}_{k}^d)^{-1}\hat{\vect{d}}_k$ with $\chi^2_{df}(\alpha)$:
\begin{equation}
\begin{aligned}
&{\rm Accept \ H_0 \ if \ } \hat{\vect{d}}_k^\top (\vect{P}_{k}^d)^{-1}\hat{\vect{d}}_k \leq \chi^2_{df}(\alpha)\\
&{\rm Accept \ H_1 \ if \ } \hat{\vect{d}}_k^\top (\vect{P}_{k}^d)^{-1}\hat{\vect{d}}_k > \chi^2_{df}(\alpha),
\end{aligned}\label{e003}    
\end{equation}
where $\vect{P}_{k}^d \triangleq {\mathbb E}[\tilde{\vect{d}}_k\tilde{\vect{d}}_k^\top]=\vect{C}^G(\vect{A} \vect{P}_{k-1}\vect{A}^\top+\vect{\Sigma}_w)(\vect{C}^G)^\top+\vect{\Sigma}_G$, and $\chi_{df}^2(\alpha)$ is the threshold found in the Chi-square table. In $\chi_{df}^2(\alpha)$, $df$ denotes the degree of freedom, and $\alpha$ denotes the statistical significance level.

The proposed $\chi^2$ CUSUM detector is characterized by the detector state $S_k\in\mathbb{R}$:
\begin{align}
    S_{k}=\delta S_{k-1}+(\hat{\vect{d}}_k)^\top (\vect{P}_{k}^d)^{-1}\hat{\vect{d}}_k, \quad S_0=0,
    \label{e003.1}
\end{align}
where $0<\delta<1$ is the pre-determined forgetting factor. The attack detector will raise an alarm
\begin{equation*}
    \text{if}\quad S_k>\sum_{i=0}^{\infty}\delta^i\chi^2_{df}(\alpha)=\frac{\chi^2_{df}(\alpha)}{1-\delta}.
\end{equation*}
\begin{remark}
Comparing to standard CUSUM algorithm in~\cite{page1954continuous}, the proposed CUSUM detector has asymptotic behavior, where the impact of the attacks on the detector state $S_k$ decays asymptotically.
\end{remark}

\subsection{Emergency mode}\label{sec:emergency-mode}
When an attack is detected, the defender switches emergency mode on.
Let us denote $k^a$ the time when the attack is detected, which satisfies $S_{k^a}>\frac{\chi^2_{df}(\alpha)}{1-\delta}$. As well as normal mode, the state estimation and attack detection continue in normal mode. However, the state estimation can only use the output measured by IMU in emergency mode. Without GPS output, IMU based estimation accumulates error and eventually diverges as we analyze in the following section.

\textbf{State estimation.}
The state estimation algorithm~\eqref{eq:P_update} with $\vect{K}_k^G=0$ is used to recursively estimate the state $\hat{\vect{x}}_k$ and the error covariance $\vect{P}_k$.

\textbf{Attack detection.} At each time $k$, the CUSUM detector
~\eqref{e003.1} is used to update the detector state $S_k$ and detect the attack.
The corresponding covariance can be found by
$\vect{P}_{k}^d \triangleq {\mathbb E}[\hat{\vect{d}}_k\hat{\vect{d}}_k^\top]=\vect{C}^G(\vect{A} \vect{P}_{k-1}\vect{A}^\top+\vect{\Sigma}_w)(\vect{C}^G)^\top+\vect{\Sigma}_G$. If $S_k<\frac{\chi^2_{df}(\alpha)}{1-\delta}$, then it returns to the normal mode. 

\section{Stability and escape time}\label{sec:esc}
This section presents analysis on stability of the state estimation and estimation error escape time.

\subsection{Stability analysis of state estimation}
In this section, we would like to discuss stability and instability conditions of the proposed estimator. Toward this end, we show that the estimator is stable in the normal mode, and unstable in the emergency mode.

In particular, the state estimate $\hat{\vect{x}}_k$ obtained through observer~\eqref{eq:P_update} is unbiased.
Moreover, its covariance is bounded, if $(\vect{C}^G,\vect{A})$ is detectable, as shown in the following theorem. 

\begin{theorem}
Given ${\mathbb E}[\hat{\vect{x}}_0]=x_0$, we have ${\mathbb E}[\hat{\vect{x}}_k]=\vect{x}_k$ for all $k\geq0$.
If $(\vect{C}^G,\vect{A})$ is detectable, then $\vect{P}_k$ is bounded.
\label{the1.0}
\end{theorem}
\begin{pf}
State estimation error $\tilde{x}_k=x_k-\hat{x}_k$ can be described by
\begin{equation}
\begin{aligned}
\tilde{\vect{x}}_k &= \vect{A}\tilde{\vect{x}}_{k-1}+\vect{K}_k^G(\vect{A}\tilde{\vect{x}}_{k-1}+\vect{w}_{k-1}+\vect{v}_k^G)\nnum\\
&+\vect{K}_k^I(\vect{C}^I(\vect{A}\tilde{\vect{x}}_{k-1}-\tilde{\vect{x}}_{k-1}+\vect{w}_{k-1})+\vect{v}_k^I).
\end{aligned}\label{e002.1}
\end{equation}

Notice that ${\mathbb E}[\vect{w}_k]=0$, ${\mathbb E}[\vect{v}_k^G]=0$, and ${\mathbb E}[\vect{v}_k^I]=0$.
Given ${\mathbb E}[\hat{\vect{x}}_0]=\vect{x}_0$, we have ${\mathbb E}[\hat{\vect{x}}_1]=\vect{x}_1$ by~\eqref{e002.1}.
Assume ${\mathbb E}[\hat{\vect{x}}_{k-1}]=\vect{x}_{k-1}$, then ${\mathbb E}[\hat{\vect{x}}_k]=\vect{x}_k$ by~\eqref{e002.1}.
By induction, we conclude that ${\mathbb E}[\hat{\vect{x}}_k]=\vect{x}_k$ for $\forall k$. Therefore, ${\mathbb E}[\hat{\vect{x}}_k]=\vect{x}_k$.

Consider $\vect{P}_k(\vect{K}_k^G,\vect{K}_k^I)$, where we emphasize the dependency of $\vect{P}_k$ on $\vect{K}_k^G$, $\vect{K}_k^I$.
With $\vect{K}_k^I=0$, $\vect{P}_k(\vect{K}_k^G,0)$ is bounded by Corollary 5.2 in~\cite{anderson1981detectability}, because $(\vect{C}^G,\vect{A})$ is detectable.
Since $\vect{K}_k$ is chosen optimal, we have $\vect{P}_k(\vect{K}_k^G,\vect{K}_k^I)\leq \vect{P}_k(\vect{K}_k^G,0)$. This completes the proof. \oprocend
\end{pf}
If the GPS signals are not available for state estimation, the covariance of the state estimation is expected to be unstable over time under Assumption~\ref{asm1}. Let us first define necessary notations: $\bar{\vect{C}}^I \triangleq \vect{C}^I(\vect{I}-\vect{A}^{-1})$, $\bar{\vect{A}} \triangleq (\vect{I}-\vect{L}\bar{\vect{C}}^I)\vect{A}$ and $\vect{L} \triangleq \vect{\Sigma}_w (\vect{C}^I\vect{A}^{-1})^\top((\bar{\vect{C}}^I+\bar{\vect{C}}^I\vect{A}^{-1})\vect{\Sigma}_w (\vect{C}^I\vect{A}^{-1})^\top+\vect{\Sigma}_I)^{-1}$.

\begin{assumption} \label{asm1}
     Matrix $\vect{A}$ is invertible, and the pair $(\bar{\vect{C}}^I, \bar{\vect{A}})$ is not detectable.
\end{assumption}

\begin{remark}
Assumption~\ref{asm1} is satisfied with the double integrator model with IMU, which is widely used to design navigation controller for UAVs as in~\cite{kerns2014unmanned}.
In this case, $\bar{\vect{C}}^I = \vect{C}^I(\vect{I}-\vect{A}^{-1}) = 0$, and $\bar{\vect{A}} =\vect{A}$, where $\vect{A}$ is not asymptotically stable.
Then, the pair $(\bar{\vect{C}}^I, \vect{A})$ is not detectable and it fulfills Assumption~\ref{asm1}.
\label{rmk1}
\end{remark}

\begin{theorem}
Assume $\vect{K}_k^G =0$. Then, under Assumption~\ref{asm1},
$\vect{P}_k$ is increasing unboundedly.
\label{the1}
\end{theorem}
\begin{pf}
We prove the statement by finding an equivalent Kalman filtering problem, and then use the existing stability result for the KF in~\cite{anderson1981detectability}.

Given $\vect{K}_k^G =0$, the state estimation error update law can be obtained from~\eqref{eq:P_update}:
\begin{align*}
\tilde{\vect{x}}_{k} &= \vect{A} \tilde{\vect{x}}_{k-1} - \vect{K}_k^I (\vect{C}^I(\vect{I}-\vect{A}^{-1})\vect{A} \tilde{\vect{x}}_{k-1}+\vect{C}^I\vect{w}_{k-1}+\vect{v}_k^I)\nnum\\
&= \vect{A} \tilde{\vect{x}}_{k-1} - \vect{K}_k^I (\vect{C}^I(\vect{I}-\vect{A}^{-1})\vect{A} \tilde{\vect{x}}_{k-1}\nnum\\
&+\vect{C}^I (\vect{I}-\vect{A}^{-1})\vect{w}_{k-1}+\vect{C}^I\vect{A}^{-1}\vect{w}_{k-1}+\vect{v}_k^I).
\end{align*}
The above state estimation error update is the Kalman filter solution of
\begin{align*}
    \vect{x}_{k} &= \vect{A}\vect{x}_{k-1} + \vect{w}_{k-1}\nnum\\
    \vect{y}_k&=\bar{\vect{C}}^I\vect{x}_k + \bar{\vect{v}}_k^I,
\end{align*}
where $\bar{\vect{v}}_k^I\triangleq\vect{C}^I\vect{A}^{-1}\vect{w}_{k-1}+\vect{v}_k^I$. However, the process noise and the measurement noise are coupled; i.e., ${\mathbb E}[\vect{w}_{k-1}\bar{\vect{v}}_k^\top]=\vect{\Sigma}_w(\vect{A}^{-1})^\top(\vect{C}^I)^\top \neq 0$. Then, the optimal gain is different from the Kalman gain used in the standard KF.
To decouple the noises, it is a common practice to add zero term $\vect{L}(\vect{y}_k-\bar{\vect{C}}^I\vect{x}_k - \bar{\vect{v}}_k^I)$ to the state equation above:
\begin{align*}
    \vect{x}_{k} &= \vect{A}\vect{x}_{k-1} + \vect{w}_{k-1}\\
                 &  \, + \vect{L}(\vect{y}_{k}-\bar{\vect{C}}^I(\vect{A}\vect{x}_{k-1}+\vect{w}_{k-1}) - \bar{\vect{v}}_k^I)\nnum\\
    &=(\vect{I}-\vect{L}\bar{\vect{C}}^I)\vect{A}\vect{x}_{k-1} + \vect{L}\vect{y}_k + (\vect{I}-\vect{L}\bar{\vect{C}}^I)\vect{w}_{k-1} - \vect{L}\bar{\vect{v}}_k^I\nnum\\
    &=\bar{\vect{A}}\vect{x}_k + \bar{\vect{u}}_k +\bar{\vect{w}}_k,
\end{align*}
where $\bar{\vect{u}}_k=\vect{L}\vect{y}_k$, and $\bar{\vect{w}}_{k-1}=(\vect{I}-\vect{L}\bar{\vect{C}}^I)\vect{w}_{k-1} - \vect{L}\bar{\vect{v}}_k^I$.
The gain $\vect{L}$ is chosen such that the process noise and measurement noise are decoupled:
\begin{align*}
    {\mathbb E}[\bar{\vect{w}}_k\bar{\vect{v}}_k^\top]&=
    (\vect{I}-\vect{L}\bar{\vect{C}}^I-\vect{L}\bar{\vect{C}}^I\vect{A}^{-1})\vect{\Sigma}_w (\vect{C}^I\vect{A}^{-1})^\top-\vect{L}\vect{\Sigma}_I\nnum\\
    &=0,
\end{align*}
where the solution is
{\small
\begin{align*}
    \vect{L}=\vect{\Sigma}_w (\vect{C}^I\vect{A}^{-1})^\top((\bar{\vect{C}}^I+\bar{\vect{C}}^I\vect{A}^{-1})\vect{\Sigma}_w (\vect{C}^I\vect{A}^{-1})^\top+\vect{\Sigma}_I)^{-1}.
\end{align*}
}
Now, the state estimation error update law~\eqref{eq:P_update} and its covariance update law~\eqref{eq:P_update} with $\vect{K}_k^G=0$ are the Kalman filtering solution of
\begin{align*}
    \vect{x}_{k+1} &= \bar{\vect{A}}\vect{x}_k +\bar{\vect{u}}_k+ \bar{\vect{w}}_k\nnum\\
    \vect{y}_k&=\bar{\vect{C}}^I\vect{x}_k + \bar{\vect{v}}_k^I,
\end{align*}
where the process noise and the measurement noise are decoupled from each other.
Under Assumption~\ref{asm1}, $\vect{P}_k$ is unstable by Corollary 5.2 in~\cite{anderson1981detectability}. This completes the proof. \oprocend
\end{pf}
\begin{remark}
Stability with GPS (Theorem~\ref{the1.0}) and instability without GPS (Theorems~\ref{the1}) are generalized into a system with any relative sensors.
This is because output model $\vect{y}_k^I$ in~\eqref{e000} can represent any relative measurement sensor model regardless of their internal error state dynamic.
This is a sharp contrast to existing analytical result. For example, it is a common practice to introduce error state variables for IMU as in~\cite{bevly2007cascaded}. Using their specific system and output model, one can check detectability of the augmented system to guarantee stability/instability. However, it is not generalized because each system has different system/output(sensor) matrices.
\oprocend \end{remark}

\subsection{Escape time of the state estimation error}
This section proposes a new resilience measure, the escape time, and analyzes the escape time in the emergency mode.

It has been revealed in Theorem~\ref{the1} that the state estimation becomes less trustful, if GPS signals are compromised for a long time. Therefore, the UAV should escape from the GPS spoofer at a certain time before the estimation becomes unreliable. Formally, we define the escape time as follows.
\begin{definition}
The escape time $k^{esc} \geq 0$ is
the time difference between the attack time $k^a$ and the first time instance when the estimation error $\vect{x}_k-\hat{\vect{x}}_k$ may not be in tolerable error distance $\zeta \in {\mathbb R}^n$ with the significance $\alpha$, i.e.
\begin{align*}
    k^{esc}=&\argmin_{k \geq k^a} k - k_a\nnum\\
    &{\rm s.t. \ } \zeta^\top \vect{P}_{k}^{-1}\zeta < \chi_{df}^2(\alpha).
\end{align*}
\label{def1}
\end{definition}
Given the desired confidence level $\alpha$, the degree of freedom $df$, we have $\chi_{df}^2(\alpha)$.
Since the optimal gains $\textbf{K}_{k^a}$, $\textbf{K}_{k^a+1}$, $\cdots$ can be found in advance at $k=k^a$, the corresponding covariance matrices $\vect{P}_{k^a},\vect{P}_{k^a+1},\cdots$ can also be found by the covariance update law in~\eqref{eq:P_update}.
Algorithm~\ref{algo1} presents escape time calculation.
Given $k^a$, the state estimation errors may not remain in the tolerable region with the predetermined confidence $\alpha$ after $k^{esc}$.
\begin{algorithm}[!ht]
\caption{Escape time calculation}
\algorithmicrequire $k^a$, $\alpha$, $df$, $\zeta$; \\
\algorithmicensure $k^{esc}$;
\begin{algorithmic}[1]
\State $k=k^a$;
\WHILE{$\zeta^T\vect{P}_k^{-1}\zeta > \chi_{df}^2 (\alpha)$}
\State $k=k+1$;
\ENDWHILE
\State $k^{esc} = k - k^a$.
\end{algorithmic}\label{algo1}
\end{algorithm}

If $\bar{\vect{C}}^I=0$ as in Remark~\ref{rmk1}, then a lower bound of the escape time can also be found before actually operating the UAV as shown in Theorem~\ref{the2}.
Let $\vect{P}$ denote the stationary point of the covariance update~\eqref{eq:P_update}, i.e.
\begin{equation*}
        \vect{P}=f\left(\vect{P}, g(\vect{P})\right),
\end{equation*}
where $f(\cdot)$ and $g(\cdot)$ were defined in~\eqref{eq:P_update} and~\eqref{eq:K_law} respectively.
Matrices used in the theorem are defined by
\begin{align*}
    \bar{\vect{\Sigma}} &\triangleq (\vect{I}-\vect{K}_k^I\vect{C}^I)\vect{\Sigma}_w (\vect{I}-\vect{K}_k^I\vect{C}^I)^\top+ \vect{K}_k^I \vect{\Sigma}_I (\vect{K}_k^I)^\top.
\end{align*}

\begin{theorem}\label{the2}
Assume $\bar{\vect{C}}^I=0$ and GPS spoofing attacks start after $\vect{P}_k$ converges to $\vect{P}$.
Then, a lower bound of the escape time can be found by 
\begin{itemize}
\item For $\|\vect{A}\|\neq1$,
    \begin{align*}
    k^{esc}
    \geq \left(\frac{1}{\log \|\vect{A}\|^2}\right)
    \log
    \left(
    \frac
    {\frac{\|\zeta\|^2}{\chi_{df}^2(\alpha)}
    +
    \frac{\|\bar{\vect{\Sigma}}\|}{\|\vect{A}\|^2-1}}
    {\|\vect{P}\|
    +
    \frac{\|\bar{\vect{\Sigma}}\|}{\|\vect{A}\|^2-1}}
    \right)
    \end{align*}
\item For $\|\vect{A}\|=1$,
    \begin{align*}
         k^{esc} \geq (\|\zeta\|^2/\chi_{df}^2(\alpha) - \|\vect{P}\|)/ \|\bar{\vect{\Sigma}}\|.
    \end{align*}
\end{itemize}
\end{theorem}
\begin{pf}
First, we derive an upper bound on $\| \vect{P}_k \|$. Assume $k^a=0$ and $\vect{P}_{0}=\vect{P}$ without loss of generality. Given $\vect{K}_k^G =0$, the covariance update law can be obtained from~\eqref{eq:P_update}:
\begin{equation}\label{eq:P_update_wo_GPS_short}
\vect{P}_k= \vect{A}\vect{P}_{k-1}\vect{A}^\top + \bar{\vect{\Sigma}}, \quad \vect{P}_0=\vect{P},
\end{equation}
where $\bar{\vect{\Sigma}}$ is time-invariant because the optimal $\vect{K}_k^I$ is time-invariant:
\begin{align*}
\vect{K}_k^I &= \vect{\Sigma}_w(\vect{C}^I)^\top (\vect{C}^I\vect{\Sigma}_w(\vect{C}^I)^\top+\vect{\Sigma}_I)^{-1}.
\end{align*}
Applying basic matrix norm property to~\eqref{eq:P_update_wo_GPS_short}, we have
\begin{equation*}
    \|\vect{P}_k\| \leq \|\vect{A}\|^2\| \vect{P}_{k-1}\|
    + \|\bar{\vect{\Sigma}}\|, \quad \|\vect{P}_0\|=\|\vect{P}\|.
\end{equation*}
By recursively applying the bound above, we have
\begin{equation}\label{eq:upperbnd}
\begin{aligned}
    \|\vect{P}_k\| &\leq \|\vect{A}\|^{2k}\|\vect{P}_{0}\| + \|\bar{\vect{\Sigma}}\|\sum_{i=0}^{k-1}\|\vect{A}\|^{2i}\\
\end{aligned}
\end{equation}
Now, we will apply the bound~\eqref{eq:upperbnd} to the $\chi^2$ test equation.
For the positive definite matrices $\vect{P}_k$, we have
\begin{align*}
    \|\vect{P}_k\|^{-1}\|\zeta\|^2 = \lambda_{\min}(\vect{P}_k^{-1})\|\zeta\|^2\leq \zeta^\top \vect{P}_k^{-1} \zeta,
\end{align*}
and thus the time $k_\Delta$  that verifies
\begin{align}
    (\|\vect{A}\|^{2k_\Delta}\|\vect{P}_{0}\| + \|\bar{\vect{\Sigma}}\|\sum_{i=0}^{k_\Delta-1}\|\vect{A}\|^{2i})^{-1}\|\zeta\|^2 =\chi_{df}^2(\alpha)
\label{the:eq1}
\end{align}
is a lower bound of the escape time $k_\Delta \leq k^{esc}$.
If $\|\vect{A}\|=1$, then
\begin{align*}
    \|\zeta \|^2 / \chi_{df}^2(\alpha) &= \|\vect{P}_{0}\| + k_\Delta\|\bar{\vect{\Sigma}}\|,
\end{align*}
which proves the theorem.
If $\|\vect{A}\| \neq 1$, by the sum of geometric series, equation~\eqref{the:eq1} becomes
\begin{align}
    \|\zeta \|^2 / \chi_{df}^2(\alpha) &= \|\vect{A}\|^{2k_\Delta}\|\vect{P}_{0}\| + \|\bar{\vect{\Sigma}}\|\frac{\|\vect{A}\|^{2k_\Delta} - 1}{\|\vect{A}\|^2 - 1}.
\label{eq:bnd2}
\end{align}
By taking log on both sides, we have the desired result.
\oprocend
\end{pf}

\section{Discussion}\label{sec:disc}
As a new resilience measure, the escape time provides a new criterion for optimal path planning with increasing uncertainties. In this section, we discuss relevant problems. 

\begin{figure}[thpb]
\centering
 \includegraphics[width=0.4\textwidth]{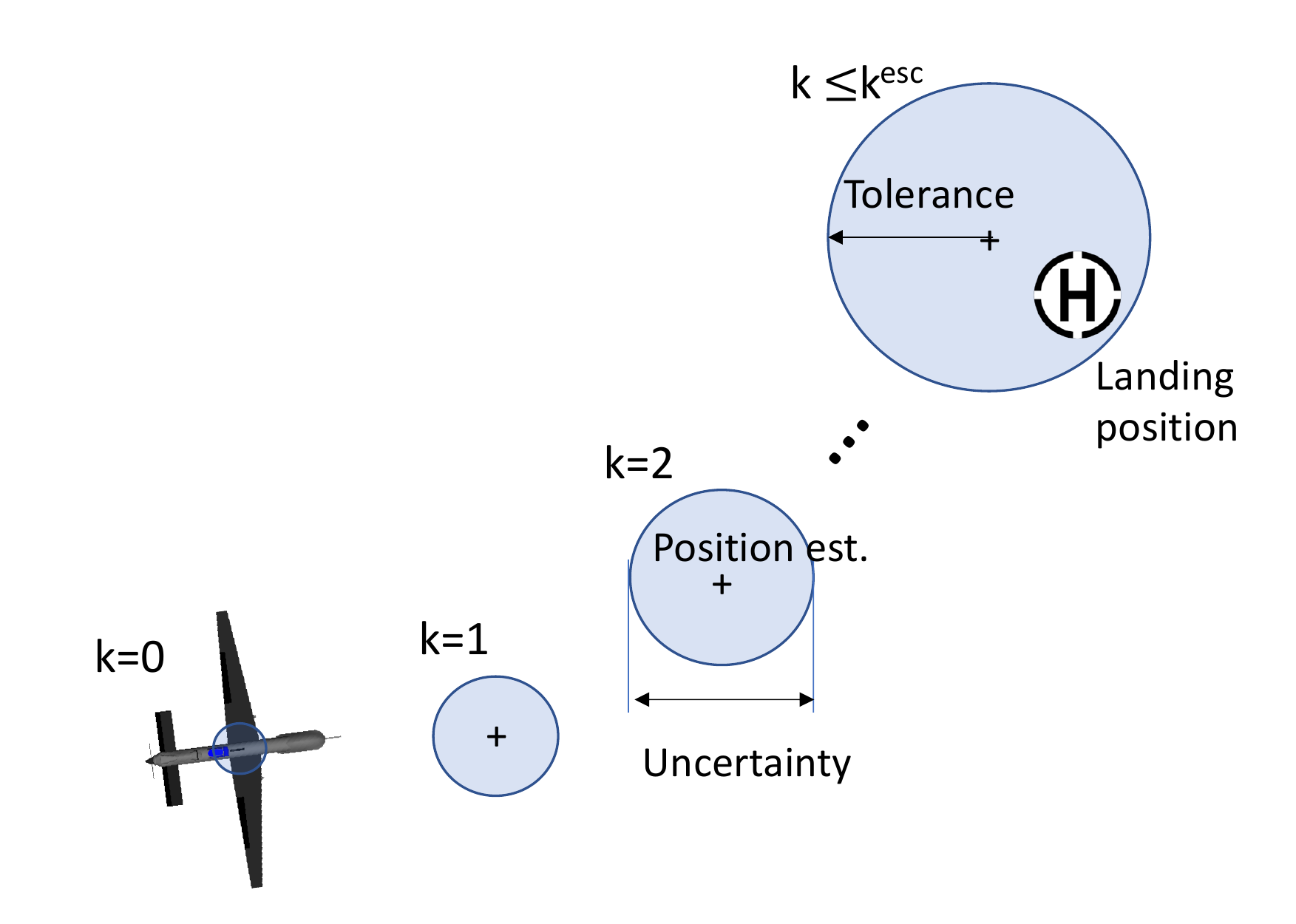}
 \caption{Illustration of path planning problem considering the increasing uncertainties}
\medskip
\label{fig:pathplaning}
\end{figure}
Once an attack is detected, the UAV is expected to land on a secure position in the pre-calculated escape time if possible.
The path planning problem illustrated in Figure~\ref{fig:pathplaning} can be formulated as
\begin{align*}
    &Optimal \ path \ planning \ problem\nnum\\
 &{\rm s.t. \ } \vect{x}_{k^{esc}} =\vect{x}_{\text{landing position}}
\end{align*}
where the UAV should arrive at the landing position before the escape time.

On the other hand, if a secure landing position is not available, the UAV is expected to escape from the spoofer within the escape time. If the output power of the GPS spoofing signal is time invariant, the defender is able to spot the spoofer through measurement of the GPS signal strength (e.g., signal-to-noise ratio) with corresponding state estimates. Then, the problem of interest becomes
\begin{align*}
    &Optimal \ path \ planning \ problem\nnum\\
 &{\rm s.t. \ } P(\|\vect{x}_{k^{esc}}-\vect{x}_{\text{spoofer}}\|>c_{\text{safe}})>\gamma,
\end{align*}
where $\vect{x}_{\text{spoofer}}$ is the location of the spoofer, $c_{\text{safe}}$ is a pre-determined distance such that the GPS spoofing signal cannot affect the UAV, and $\gamma<1$ denotes the desired certainty.

\section{Simulations}\label{sec:sim}
\begin{figure}[thpb]
\centering
 \includegraphics[width=0.4\textwidth]{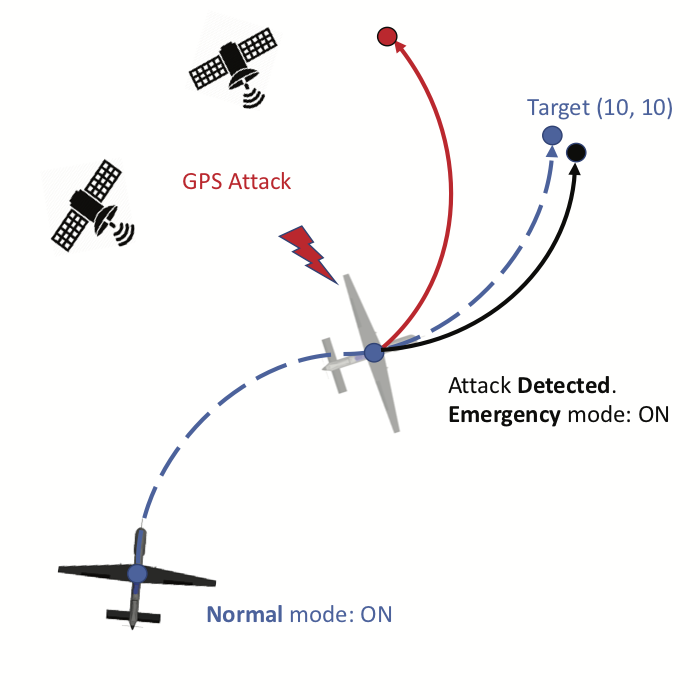}
 \caption{Illustration of the simulation scenario: (1) red line denotes the flight path with normal mode under GPS attack; (2) black line denotes the path with emergency mode.}
\medskip
\label{fig:simulation-scenario}
\end{figure}

We simulate scenarios, where a UAV gets a GPS spoofing attack during a flight to a target position.
In the first scenario, we simulate a system without attack detector, where the UAV keeps using the normal state estimator in Section ~\ref{sec:normal-mode} during the flight.
In the second scenario, the UAV detects the attack and then switches to the emergency mode as illustrated in Figure~\ref{fig:simulation-scenario}.

We use a double integrator UAV dynamics under the GPS spoofing attack as in~\cite{kerns2014unmanned}. The discrete time state vector $\vect{x}_k$ considers planar position and velocity at time step $k$, i.e.
\begin{equation*}
    \vect{x}_k = [r_k^x, r_k^y, v_k^x, v_k^y]^\top,
\end{equation*}
where $r_k^x, r_k^y$ denote x, y position coordinates, and $v_k^x, v_k^y$ denote velocity coordinates. With sampling time at $0.01$ seconds, the double integrator model is discretized into the following matrices:
\begin{equation*}
    \vect{A} = 
    \begin{bmatrix} 
    1 & 0 & 0.01 & 0    \\
    0 & 1 & 0    & 0.01 \\
    0 & 0 & 1    & 0 \\
    0 & 0 & 0    & 1
    \end{bmatrix},
    \quad
    \vect{B} = 
    \begin{bmatrix} 
    0    & 0 \\
    0    & 0 \\
    0.01 & 0 \\
    0    & 0.01 
    \end{bmatrix},
\end{equation*}
and outputs $\vect{y}^G$ and $\vect{y}^I$ measure positions from GPS and IMU, respectively, with the output matrices:
\begin{equation*}
    \vect{C}^G = 
    \begin{bmatrix} 
    1 & 0 & 0 & 0    \\
    0 & 1 & 0 & 0
    \end{bmatrix},
    \quad
    \vect{C}^I = 
    \begin{bmatrix} 
    0 & 0 & 1 & 0 \\
    0 & 0 & 0 & 1 
    \end{bmatrix}.
\end{equation*}
The covariance matrices of the sensing and disturbance noises are chosen as \begin{equation*}
    \vect{\Sigma}_w = 0.0001\vect{I}, \quad
    \vect{\Sigma}_I = 0.001\vect{I}, \quad
    \vect{\Sigma}_G = 0.001\vect{I}.
\end{equation*}
In the scenarios, the UAV is moving toward the target position with the coordinates at $(10, 10)$ by using feedback control\footnote{We implemented a proportional-derivative (PD) like tracking controller, which is widely used for double integrator systems.} using the state estimate from the state estimator. The GPS attack happens at time step $700$. The attack signal is $d = [100, 100]^\top$.

\subsection{Standard estimator (without attack detector)}
\begin{figure}[thpb]
\centering
 \includegraphics[width=0.35\textwidth]{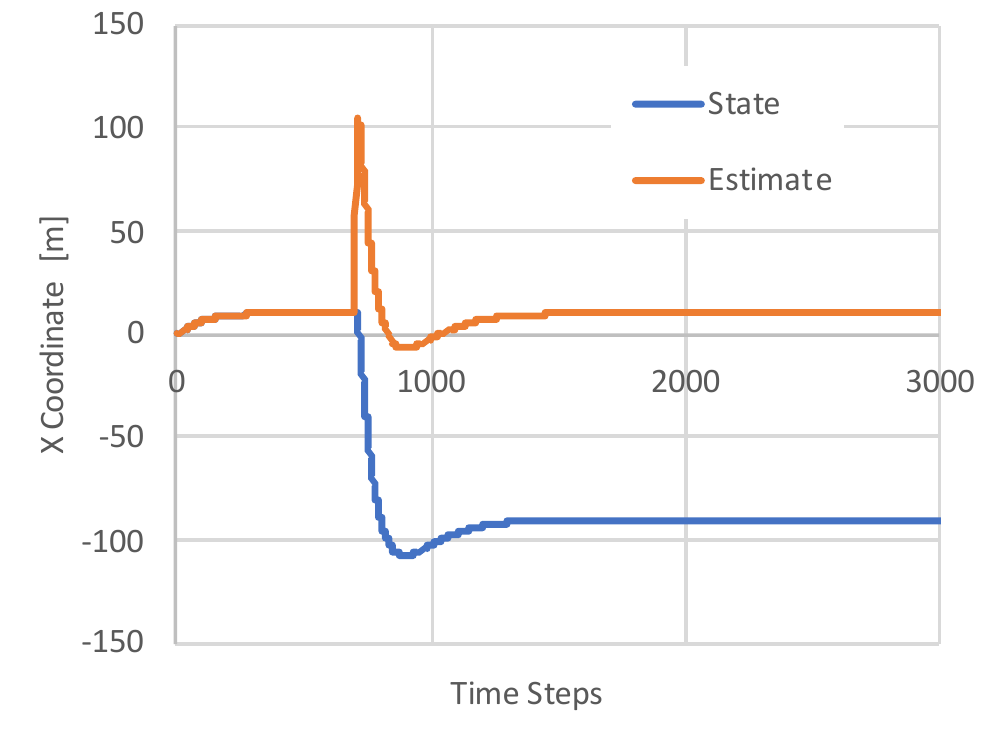}
 \caption{State and estimate under the GPS attack at $k=700$.}
 \medskip
\label{fig:2}
\end{figure}

As shown in Figure~\ref{fig:2}, the state estimation is deceived by the GPS attack. The position coordinate of the UAV actually converges toward $(-90, -90)$, however, the state estimate shows it converges to the desired position at $(10,10)$.

\subsection{Proposed method}
\begin{figure}[thpb]
\centering
\includegraphics[width=0.35\textwidth]{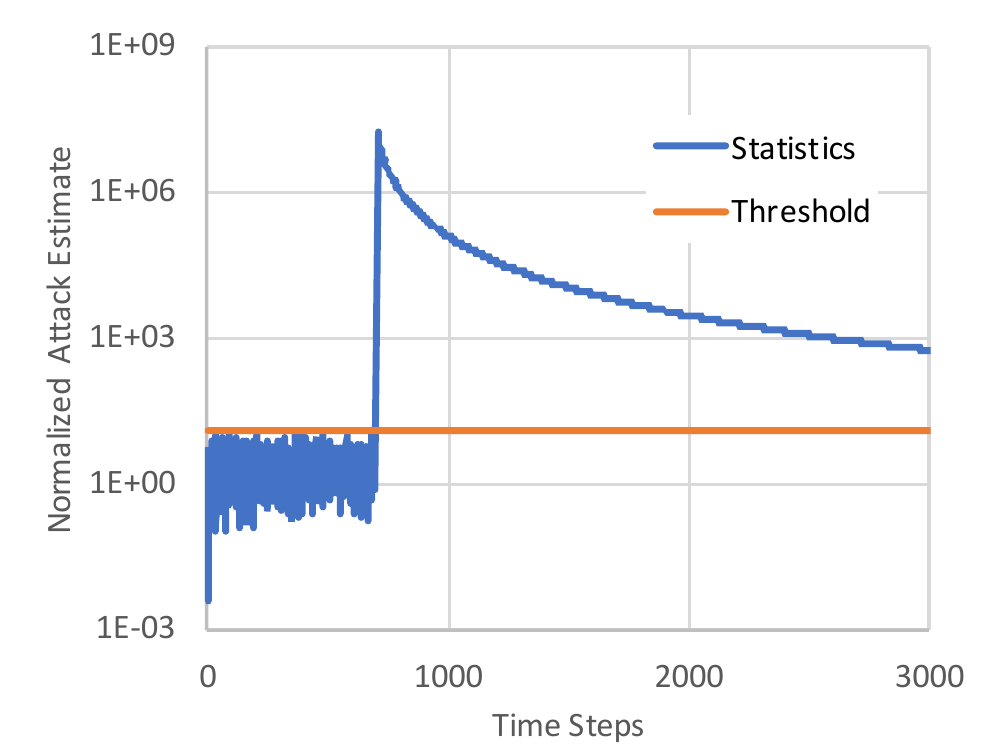}
 \caption{Attack detection: Statistics denotes $S_k$ defined in~\eqref{e003.1} of CUSUM detector. The threshold equals to $\frac{\chi_{df}^2(\alpha)}{1-\delta}$ with $\alpha=0.01$ and $\delta=0.15$.}
\medskip
\label{fig:3}
\end{figure}

The attack detector is able to detect the attack using the normalized attack vector as shown in Figure ~\ref{fig:3}. In Figure~\ref{fig:3}, there is an evident spike of the detector state, which implies there is an attack. Statistic significance of the attack is tested using the CUSUM detector described in~\eqref{e003.1} with the significance $\alpha$ at $1\%$.

Based on the hypothesis test result, we switch the system mode to the emergency.
\begin{figure}[thpb]
\centering
\includegraphics[width=0.35\textwidth]{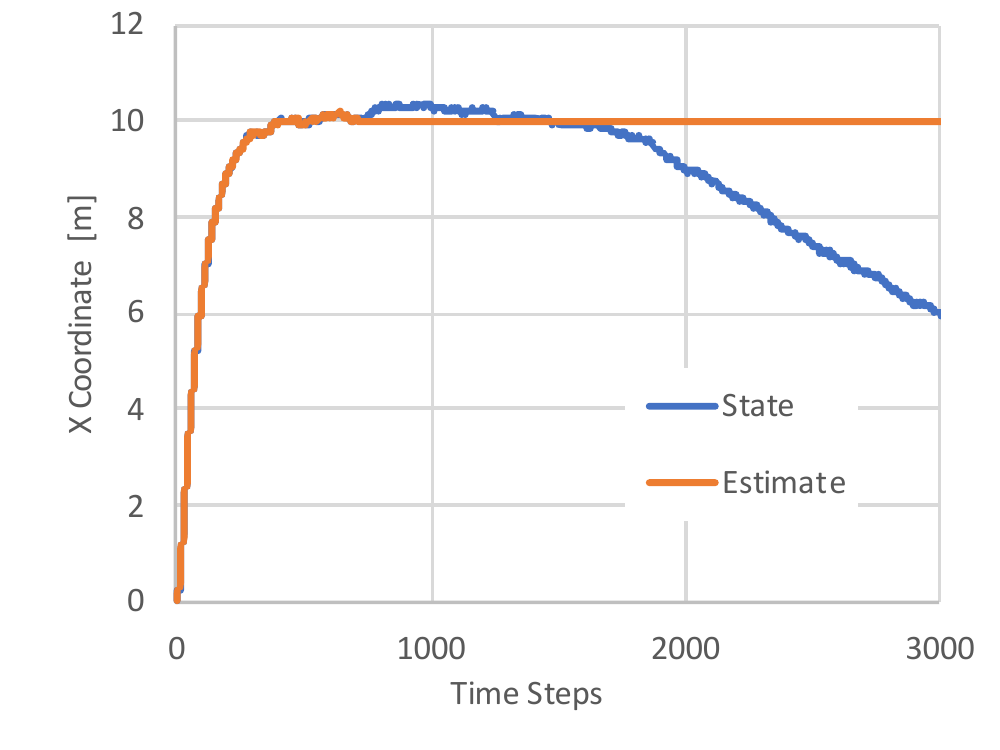}
 \caption{State and estimate with the attack mitigation.}
\medskip
\label{fig:4}
\end{figure}
As shown in Figure~\ref{fig:4}, the proposed method mitigates the attack. However, we also observe the drift of estimate in Figure~\ref{fig:4}. The drift motivates us to estimate the bounds of the drift.

\subsection{Statistical upper bound on error and escape time}
\begin{figure}[thpb]
\centering
\includegraphics[width=0.35\textwidth]{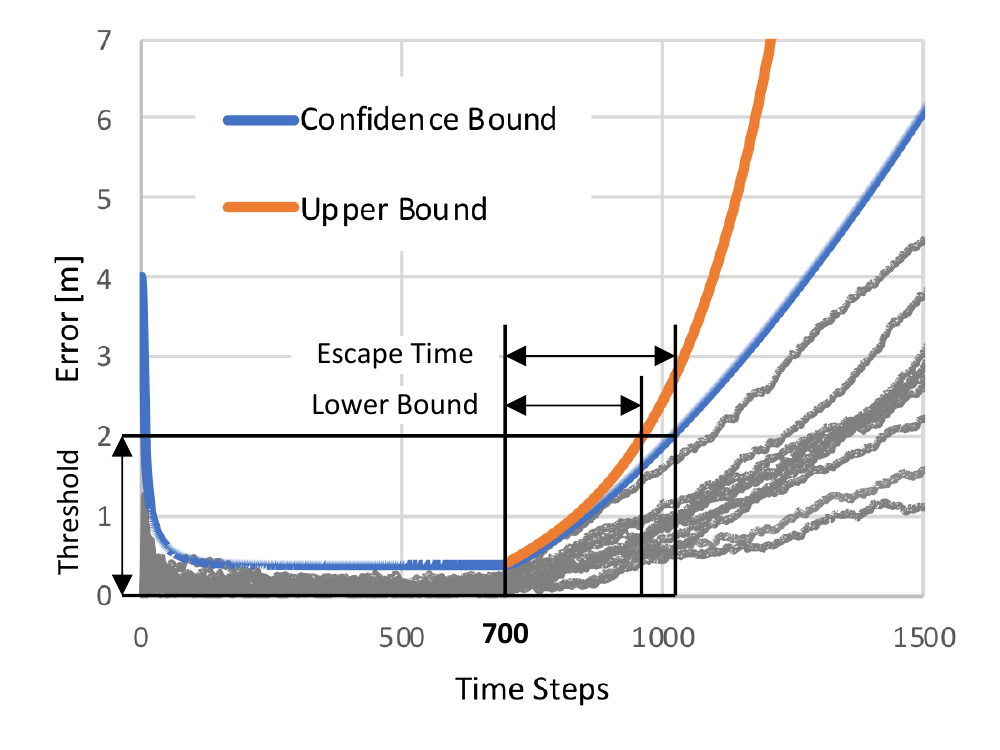}
 \caption{The $99\%$ confidence bound with the proposed method,  i. e. $\sqrt{\chi_{df}^2(\alpha) \| \vect{P}_k \| }$ with $\alpha=0.01$, the upper bound calculated in Theorem~\ref{the2}, and 10 sample trajectories (\emph{gray} colored) of $\|\vect{x}_k - \hat{\vect{x}}_k\|$.}
\medskip
\label{fig:5}
\end{figure}

Using the covariance estimate $\vect{P}_k$ and $\chi^2$ test, we can calculate a confidence interval as shown in Figure~\ref{fig:5}. The error magnitudes of 10 sample trajectories are under the $99\%$ confidence bound as shown in Figure~\ref{fig:5}. Note that the calculation of the confidence bound is deterministic. Using Algorithm~\ref{algo1}, we can calculate that it takes $290$ steps (escape time) to reach the error threshold $\|\zeta\|=2$. Alternatively, we can calculate a lower bound of the escape time according to Theorem~\ref{the2}. The lower bound is at $257$ steps which can be verified in Figure~\ref{fig:5}.

\section{Conclusion}
This paper studies resilient state estimation of UAVs in
GPS denied environment. The KF-like estimator has been designed and $\chi^2$ CUSUM algorithm is used to detect the attack. In the presence of the attack, GPS signals are not used to estimate the state, because they do not contain valid information. Due to the limited sensing device in the emergency mode, the estimation suffers from the sensor drift problem. We calculate a lower bound of the escape time, which is defined by the safe time such that the estimation error remains in a tolerable region with a high probability. A simulation of the UAV demonstrates the results.

\bibliography{references}


\end{document}